\documentclass[aps,nopacs,amsmath,amssymb,twocolumn]{revtex4}
\usepackage{graphicx}
\usepackage{epstopdf}

\usepackage{textcomp}

\def\beq{\begin{eqnarray}}
\def\eeq{\end{eqnarray}}
\def\be{\begin{equation}}
\def\ee{\end{equation}}

\def\bm{\begin{math}}
\def\me{\end{math}}

\def\q{\quad}

\def\D{\Delta}
\def\bel{\begin{equation} \label}
\def\beel{\begin{eqnarray} \label}

\newcommand \bei {\begin{itemize}}
\newcommand \eei  {\end{itemize}}

\begin{document}
\bibliographystyle{apsrev}

\title{The role of energy in ballistic agglomeration}


\author{N. V. Brilliantov$^{1}$}
\author{A. I.~Osinsky$^{1}$}
\author{P. L. Krapivsky$^{2}$}
\affiliation{$^{1}$Skolkovo Institute of Science and Technology, Moscow, Russia}
\affiliation{$^{2}$Department of Physics, Boston University, Boston, MA 02215, USA}

\begin{abstract}
We study a ballistic agglomeration process in the reaction-controlled limit.  Cluster densities obey an infinite set of
Smoluchowski rate equations, with rates dependent on the average particle energy. The latter is the same for all
cluster species in the reaction-controlled limit and obeys an equation depending on densities. We express the average energy through the total cluster density that allows us to reduce the governing equations to the standard Smoluchowski equations. We derive basic asymptotic behaviors and verify them
numerically. We also apply our formalism to the agglomeration of dark matter.
\end{abstract}

\maketitle

\section{Introduction}

In aggregation, clusters merge irreversibly upon collisions.  Aggregation is ubiquitous in Nature with applications
ranging from Brownian coagulation \cite{Smo17,Family1985,Ball1987,ThornPRL1994,Odriozola:2001,Odriozola:2004} and polymerization \cite{Flory} to atmospheric phenomena \cite{Hidy,Drake,Srivastava1982,Friedlander} and astrophysical systems
\cite{Saslaw,Lissauer:1993,Chokshietal:1993,DominikTilens:1997,Ossenkopf1993,SpahnAlbersetal:2004,esposito2006,BrilliantovPNAS2015,HEP1,HEP2,HEP3}. A complete description of aggregation is very complicated. A spectacular example of merging massive black holes has
been studied theoretically, numerically, and experimentally; this is a very complicated process. Still, the details of
the merging processes in ordinary phenomena like Brownian coagulation could be as complicated as in the black holes or
neutron stars merging. Moreover, the mass spectrum is very broad. Hence the merging is usually modeled just
postulating that it occurs with a certain rate depending on the parameters of the merging clusters. Clusters are also simply 
modeled by a single number, the mass of the cluster. Clusters are often built from
minimal-mass entities, the monomers. In this situation the mass spectrum $m_k = m_1 k$ is parametrized by integers
$k=1,2,\ldots$.

The transport mechanism plays a crucial role in aggregation. In earlier applications of aggregation to Brownian
coagulation, polymerization, and other physical and chemical processes, diffusion is the dominant {\it transport}  mechanism, e.g.  \cite{Family1985,Ball1987,ThornPRL1994,Odriozola:2001,Odriozola:2004}. Thus the particles have random rather than deterministic trajectories. Such aggregation processes are well understood \cite{Leyvraz2003,krapbook}. The main quantities of
interest are cluster densities $n_k(t)$ which depend only on the mass $k$ and time $t$. In the homogeneous setting, these
densities evolve according to Smoluchowski rate equations. For infinite systems, Smoluchowski's equations are an infinite
system of nonlinear coupled ordinary differential equations depending on merging rates. Smoluchowski equations have
been analytically solved only in a few cases, namely for the  general bilinear kernel \cite{Drake,bilinearkernel2}; more recently, exact solutions have been established for the parity kernel \cite{paritykernel} and the $q$-sum kernel \cite{qsumkernel}. Scaling analysis \cite{Ernst1985,Ernst1988} often provides a good qualitative understanding of the most interesting large time behavior.

Ballistic transport also underlies many aggregation processes such  as aggregation of dust in interplanetary space and
particles in planetary rings
\cite{Chokshietal:1993,DominikTilens:1997,Ossenkopf1993,SpahnAlbersetal:2004,esposito2006,BrilliantovPNAS2015}. Since
diffusive transport is usually tacitly assumed when aggregation is mentioned, we shall use the term ballistic
agglomeration (BA) to describe  aggregation processes with ballistic transport. The BA processes have 
diverse applications ranging from in-space manufacturing to the evolution of the dark matter \cite{HEP1,HEP2,HEP3}.

Despite numerous studies of the BA processes 
\cite{CarnevalePameauYoung:1990,TrizacHansen:1995,Frachebourg1999,FrachebourgPiasecki2000,LF00,BA1D,Trizac:2003,BrilliantovSpahn2006,BFP2018,Das,Das1,Mazza,Mazza1}, our understanding of such systems is much less complete than the understanding of the
diffusion-driven aggregation. The key difference of the BA from diffusion-driven aggregation is the primary role of the
kinetic energy which is partially lost in merging events. In aggregation processes, each cluster is characterized by its mass; in
the BA processes, we must also account for velocities and rely on a joint mass-velocity distribution satisfying
Boltzmann-Smoluchowski equations \cite{BrilliantovSpahn2006,BrilBodKrap2009,BrilliantovPNAS2015,BFP2018}. The Boltzmann
equation is already notoriously difficult; the Boltzmann-Smoluchowski equations form an infinite set of nonlinear
coupled integro-differential equations, each one more complicated than the Boltzmann equation. One very general
solution of the Boltzmann equation, the Maxwell distribution, describes equilibrium.  If different cluster species were
at equilibrium, then velocity distributions would be known. Temperature equilibrium (temperature equipartition) is violated for the BA: The temperatures of each species defined via the corresponding average kinetic energy are different.

Fortunately, there is a special limit when all species are close to temperature equilibrium. This is the reaction-controlled limit \cite{Trizac:2003} (see also \cite{Ball1987,ThornPRL1994,Odriozola:2001,Odriozola:2004} for the diffusive transport) when, in contrast to the collision-controlled limit, merging occurs in a tiny fraction of collisions --- clusters mostly undergo elastic collisions and therefore are near equilibrium. The entire system is then characterized by the same temperature $T(t)$; it evolves in time, manifesting the non-equilibrium nature of the process. An important feature of the reaction-controlled BA is the validity of the mean-field description; in the collision-controlled BA, the mean-field Boltzmann-Smoluchowski
fails in all spatial dimensions \cite{krapbook} although the failure becomes pronounced only at a very large time and at intermediate times the deviations are usually small.

Previous work  \cite{Trizac:2003} on the reaction-controlled BA was focused on average quantities. In this paper, we develop a general framework that allows one to determine both the mass distribution and the evolution of temperature. This framework is presented in Sect.~\ref{sec:gen}. In Sect.~\ref{sec:DM} we apply our formalism to the agglomeration of dark matter. We conclude in Sect.~\ref{sec:concl}.

\section{Rate equations for ballistic aggregation}
\label{sec:gen}

\subsection{Derivation of the rate equations}

Equations governing the dynamics of the BA are derived in the realm of the Boltzmann equation
\cite{ChapmanCowling:1970,BrilliantovPoeschelOUP} approach. The main object is $f_k(v_k,t)$,  the density of clusters
of mass $k$ and velocity ${\bf v}_k$. To illustrate the basic physics, we provide a transparent  derivation based on
the direct computation of the collision rates and energy losses. We consider diluted and spatially uniform 3D systems.
We ignore the shape of clusters and effectively assume that clusters are balls: a cluster of `size' $k$ has mass
$m_k=m_1k$ and the diameter $\sigma_k = \sigma_1 k^{1/3}$.

\begin{figure}
\centering
\includegraphics[width=4cm]{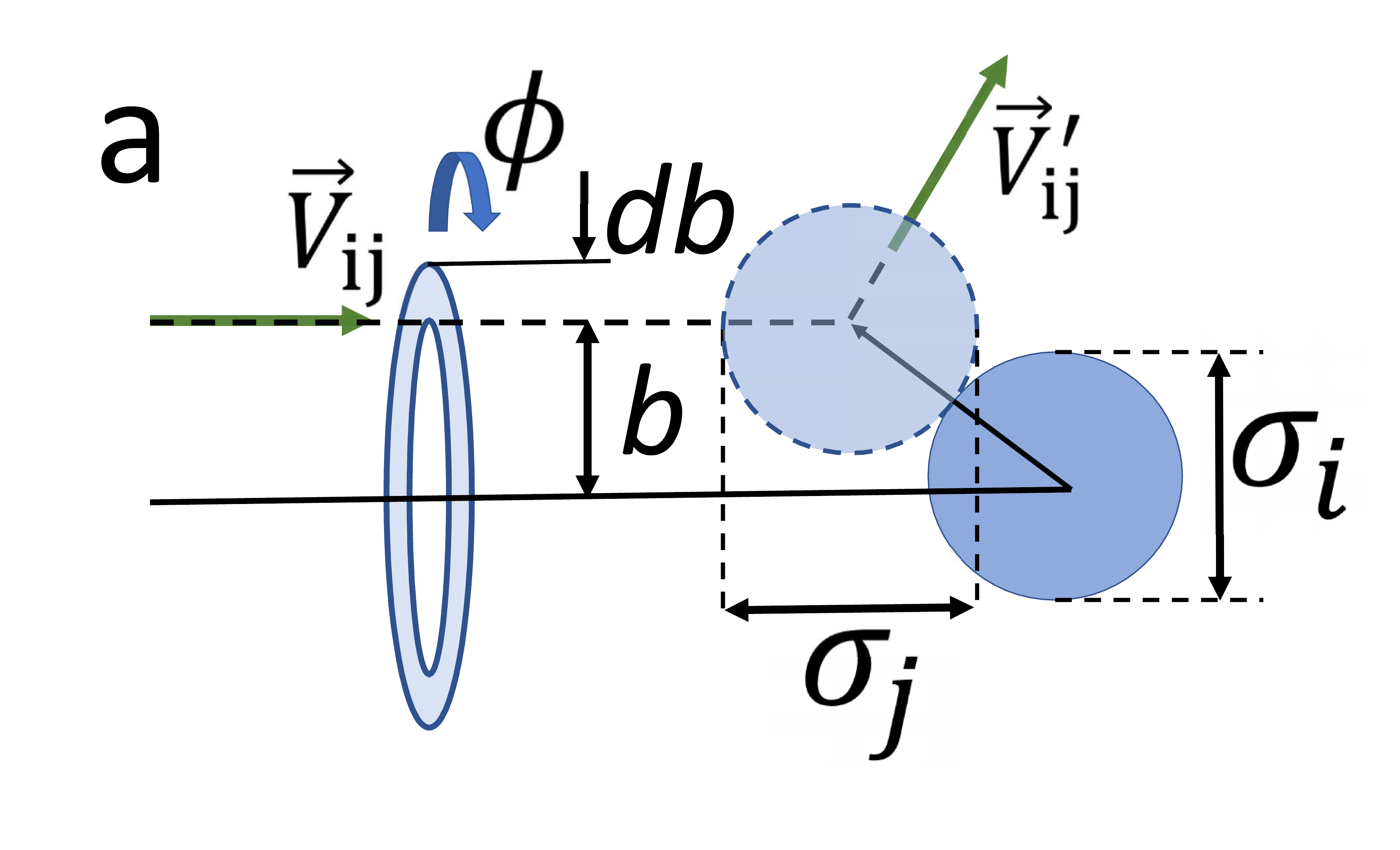}\q
\includegraphics[width=4cm]{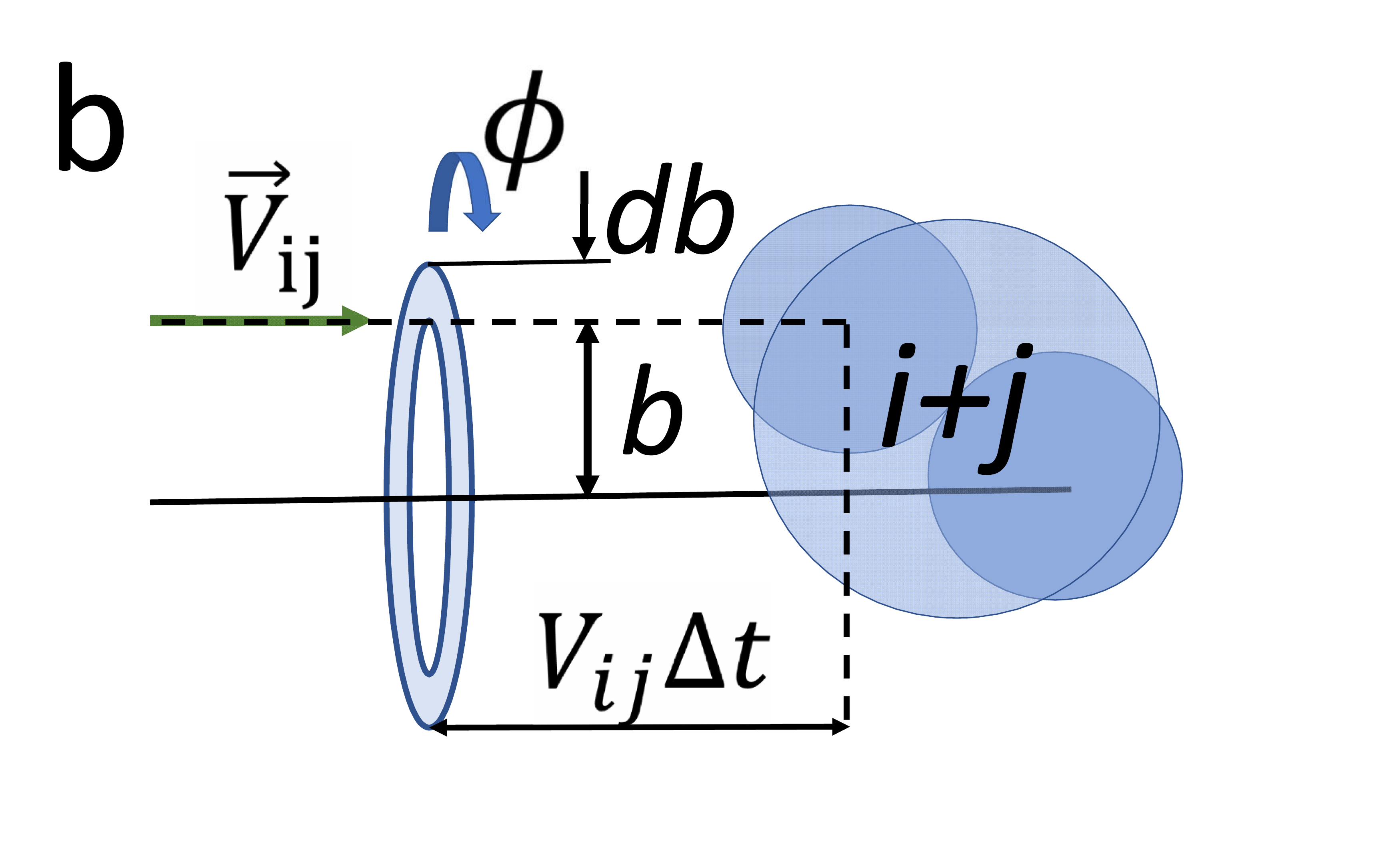}
\caption{Sketch of the bouncing (a) and aggregative impact (b) of clusters of size $i$ and $j$. The collision is
specified by the impact parameter $b$, the angle $\phi$, and the relative velocity $v_{ij}$.}
\label{Fig1}
\end{figure}

To determine the merging rate consider a collision of two clusters  with mass-velocity parameters $(i, {\bf v}_i)$ and
$(j, {\bf v}_j)$. In the coordinate system attached to $(i, {\bf v}_i)$, another cluster moves with the velocity ${\bf
v}_{ij}={\bf v}_i -{\bf v}_j$. When projected onto the plane, perpendicular to the velocity ${\bf v}_{ij}$, the position of the second cluster can be specified in the polar coordinates by the radius $b$ (the impact parameter) and the polar angle $\phi$, see Fig. \ref{Fig1}. Take clusters of mass $i$ with velocities in the tiny region of volume $d{\bf v}_i$ around
${\bf v}_i$; similarly for clusters of mass $j$. The number of collisions between such ensembles of clusters happening
during the time interval $\D t$ in a small volume $d{\bf r}$ reads
\begin{equation}
\label{Boltzmann}
f_i(v_i) d{\bf v}_i\,f_j(v_j) d{\bf v}_j\, v_{ij} \D t \,bd\phi db\,  d{\bf r}\,.
\end{equation}
The densities $f_i(v_i)$ and $f_j(v_j)$ do not depend on the spatial  location (we consider only spatially uniform
systems) and on the direction of the velocity (due to isotropy). The factor $bd\phi db v_{ij} \D t $ gives the volume
of the collision cylinder [see Fig.~\ref{Fig1}] specified by the impact parameter $b\in [b, b+db]$ and the angle $\phi$:
$bd\phi db$ is the cross-section and $v_{ij} \D t $ is the length of the cylinder. Equation \eqref{Boltzmann} is based on the assumption that the velocities of colliding clusters are uncorrelated. This assumption, first applied to molecular gases, was called a ``molecular chaos hypothesis". Here it is applied to particle physics and is expected to be accurate for diluted systems in the reaction-controlled setting. The use of the molecular chaos hypothesis in the collision-controlled setting is not completely justified.

To find the number of collisions between clusters of size $i$ and $j$ we integrate \eqref{Boltzmann} over  parameters
specifying the collision, that is, over $\phi \in [0, 2\pi]$ and $b\in [0, \sigma_{ij}]$ with $\sigma_{ij}=(\sigma_i+
\sigma_j)/2$, and also over all possible velocities ${\bf v}_i$ and ${\bf v}_j$. The agglomeration rate is therefore
\beel{2} {\cal B}_{ij} &=& \int_0^{2 \pi} \!\!d \phi \int_0^{\sigma_{ij}}bdb\int d{\bf v}_i \,f_i(v_i)
\int d{\bf v}_j \, f_j(v_j)  v_{ij}  \nonumber \\
&=& \pi \sigma_{ij}^2 \int \int d{\bf v}_i\, d{\bf v}_j\,  v_{ij} f_i(v_i) f_j(v_j) \,. \eeq
In the reaction-controlled limit, a tiny fraction of collisions leads to merging. We assume for simplicity that this fraction does not depend on the cluster size and/or on the relative velocity of the collision, although generally, this could be violated, see e.g. \cite{Odriozola:2001,Odriozola:2004}. With this assumption, we can put the fraction into the time variable to avoid cluttering the formulae.

Since almost all collisions are like in the classical gas, the velocity distribution  functions are Maxwellian:
$f_i(v_i) = n_i e^{-v_i^2/v_{0,i}^2}/(\pi^{3/2} v_{0,i}^3)$, where $n_i$ is the number density of clusters of size $i$
and $v_{0,i} = \sqrt{2T/m_i}$ is the thermal velocity of such clusters ($T$ is the temperature measured in the units of
energy; equivalently, we set the Boltzmann constant to unity).

To compute the integral in Eq.~\eqref{2} we first make the transformation,  $({\bf v}_{i}, {\bf v}_{j})\to ({\bf V},
{\bf v}_{ij})$, to the center of mass velocity ${\bf V}=(m_i {\bf v}_{i}+ m_j{\bf v}_{j})/(m_i+m_j)$ and the relative
velocity ${\bf v}_{ij} =  {\bf v}_i -{\bf v}_j$. The product of the velocity distribution functions becomes
\begin{equation*}
\label{ff}
f_i(v_i)f_j(v_j) = \frac{n_i n_j}{\pi^3 v_{0,i}^3v_{0,j}^3}\, \exp\!\left[-\frac{\mu_{ij} v_{ij}^2 + (m_i+m_j) V^2}{2T}\right]
\end{equation*}
where $\mu_{ij} = m_i m_j /(m_i+m_j)$ is the reduced mass. Inserting this expression  into \eqref{2} and using the
identity $d{\bf v}_i d{\bf v}_j = d{\bf V} d{\bf v}_{ij}$ we get a product of two Gaussian integrals. Computing the
integrals we find that the agglomeration rates are proportional to $\sqrt{T}$:
\begin{equation}
\label{BK} {\cal B}_{ij} = \sqrt{T}\,K_{ij}n_in_j\,.
\end{equation}
The mass-dependent factor of the rates is given by
\begin{equation}
\label{K:3} K_{ij} = K_0(i^{1/3} + j^{1/3})^2   \sqrt{i^{-1} +j^{-1}},
\end{equation}
where $K_{0} = \sigma_1^2 \sqrt{\pi/(2m_1)}$; see \cite{BrilBodKrap2009,BrilliantovPNAS2015,BFP2018}  for details of
such calculations. The governing equations for the densities are the Smoluchowski equations
\begin{equation}
\label{Sm1} \frac{d n_k}{dt} =  T^{1/2}\left[\frac{1}{2} \sum_{i+j=k} K_{ij}n_in_j - n_k\sum_{i\geq 1} K_{ki}
n_i\right]
\end{equation}
with a temperature-dependent factor.

Next, we derive the evolution equation for the total kinetic energy density, $\frac{3}{2}nT$, where $n= \sum_{k\geq 1} n_k$ is the total cluster density. In a collision between clusters $i$ and $j$ leading to merging, the total energy of the pair is reduced by the energy of the relative motion of the pair, $\mu_{ij} v_{ij}^2/2$. We treat the merged cluster as a single entity and thus do not account for the kinetic energy of the inner motion, which remains after the collision. To obtain the rate equation for the decay of the energy $\frac{3}{2}nT$,  we multiply the
integrand in Eq.~\eqref{2} by $\mu_{ij} v_{ij}^2/2$,  integrate over all possible velocities ${\bf v}_i$ and ${\bf
v}_j$, and sum over all $i$ and $j$. This gives the energy equation
\begin{equation}
\label{energy} \frac{d}{dt}  nT = -\frac{2}{3}\, T^{3/2}\sum_{i\geq 1} \sum_{j\geq 1}   K_{ij} n_in_j\,.
\end{equation}

We ignore the energy loss in the bouncing collisions. Hence these elastic collisions do not contribute to the evolution of the kinetic energy in \eqref{energy}. The generalization for inelastic collisions (as in granular gases \cite{BrilliantovPoeschelOUP}) is straightforward but  would complicate the notations.

\subsection{Analysis of the rate equations}

Summing Eqs.~\eqref{Sm1} yields
\begin{equation}
\label{n-eq} \frac{dn}{dt} = -\frac{1}{2}\,  T^{1/2}  \sum_{i\geq 1} \sum_{j\geq 1}   K_{ij} n_in_j\,.
\end{equation}
Massaging \eqref{energy}  and \eqref{n-eq} we obtain a neat result
\begin{equation}
\label{Tn-eq}
\frac{dT}{dn} = \frac{1}{3}\,\frac{T}{n}
\end{equation}
implying that the temperature is a purely algebraic function of the total density:
\begin{equation}
\label{Tn-sol} T(t)/T(0) = \left[n(t)/n(0)\right]^{1/3}\,.
\end{equation}
We emphasize that Eq. \eqref{Tn-eq} holds independently of such details as the shape of clusters or the fraction of the aggregation events (which can also depend on cluster sizes). However, one still needs to assume a complete elasticity of the bouncing collisions and that the fraction of merging events does not depend on the collision speeds.

We can absorb the factor $\sqrt{T}$ in Eqs.~\eqref{Sm1} into the time variable by introducing the modified time
\begin{equation}
\label{tau:def} \tau = \int_0^t dt'\, \sqrt{T(t')}\,.
\end{equation}
The corresponding Smoluchowski equations
\begin{equation}
\label{Smol-eq}
\frac{d n_k}{d \tau} = \frac12 \sum_{i+j=k} K_{ij}n_in_j - n_k\sum_{i\geq 1} K_{ki} n_i
\end{equation}
with rates \eqref{K:3} are analytically intractable.  Fortunately,  the rates \eqref{K:3} are homogeneous, namely they
satisfy
\begin{equation}
\label{hom}
K_{si, sj} = s^{\lambda} K_{i,j}\,.
\end{equation}
For rates \eqref{K:3}, the homogeneity index is $\lambda =1/6$.  The scaling approach \cite{Ernst1985,krapbook} tells
us that the total density decays as $n\sim \tau^{-1/(1-\lambda)}$, so in the present case $n\sim \tau^{-6/5}$. Using
this asymptotic together with \eqref{Tn-sol}--\eqref{tau:def} we obtain
\begin{equation*}
t \sim \int_0^\tau d\tau'\,[T(\tau')]^{-1/2} \sim \int_0^\tau d\tau'\,[n(\tau')]^{-1/6} \sim \tau^{6/5}
\end{equation*}
implying that
\begin{equation}
\label{nT:3d} n\sim t^{-1}\,, \quad \qquad T\sim t^{-1/3}\,.
\end{equation}

\begin{figure}
\centering
\includegraphics[width=8.3cm]{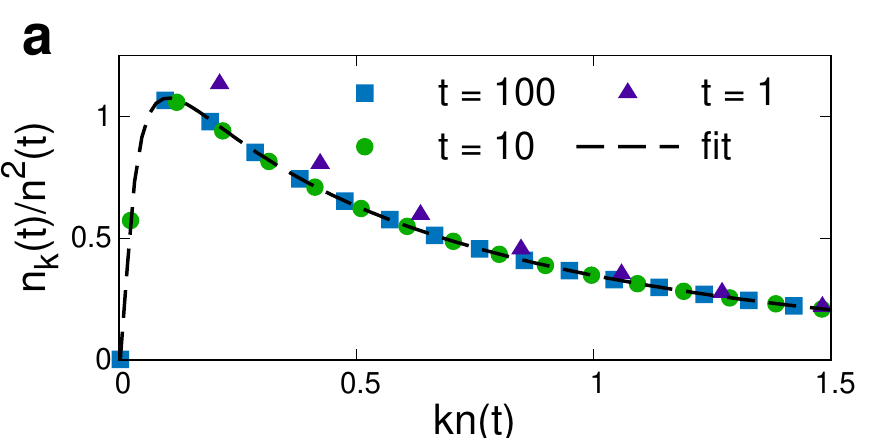}\\
\includegraphics[width=4cm]{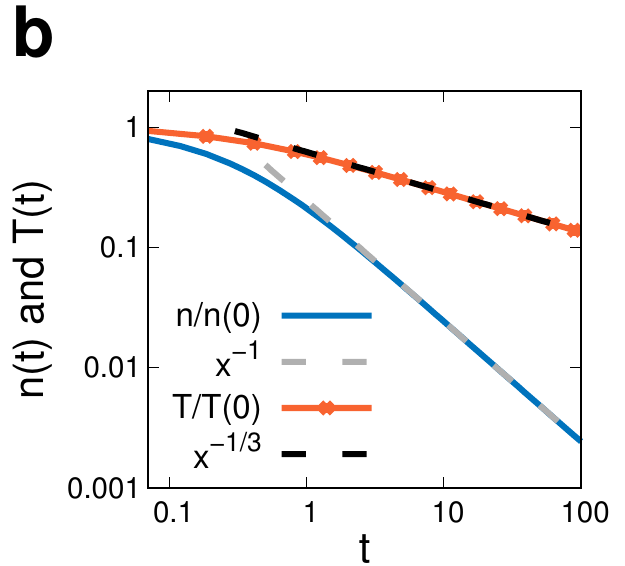}\q
\includegraphics[width=4cm]{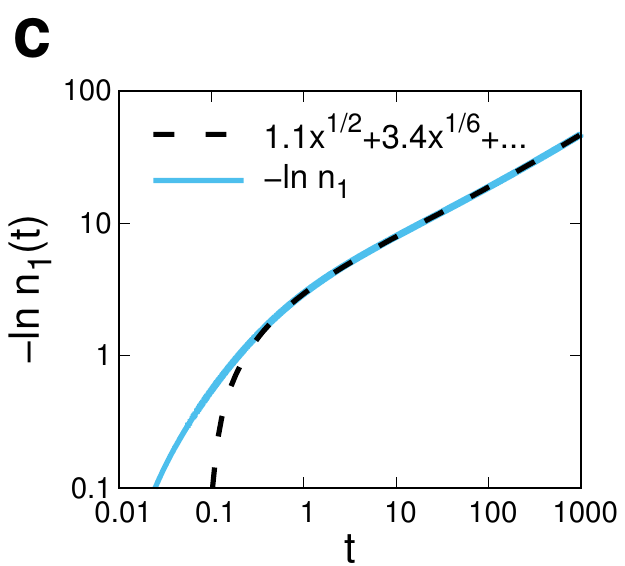}
\caption{(a) The reduced density distribution $n_k(t)/n^2(t)$ as a function of $kn(t)$ for different times for the 3D
system ($n(0)=T(0)=K_0=1$). The convergence to the scaled distribution $\Phi(x)$ with $x=kn(t)$ is observed already at
$t = 10$. The scaled distribution is well fit by $\left(\frac{0.785}{x^{0.28}} +
\frac{0.65}{x}\right)e^{-0.61/\sqrt{x}-0.81x}$. (b,c) The time dependence of temperature, the total density and monomer
density. Solid lines are numerical results; dashed lines are the asymptotic predictions, Eqs.~\eqref{nT:3d} and
\eqref{n1:3d}. The data presented in (a) and (c) are obtained by numerically solving ordinary differential equations (ODEs). The data in (b) are from Monte-Carlo simulations (see Appendix \ref{ap:num} for detail).}
 \label{Fig2}
\end{figure}

The scaling approach further predicts \cite{Ernst1985,Leyvraz2003,krapbook} that the cluster-mass distribution
approaches a scaling form
\begin{equation}
\label{scaling} n_k=n^2 \Phi(kn)
\end{equation}
in the scaling limit $t\to\infty$ ($\tau \to \infty$), $k\to\infty, ~kn=\text{finite}$. Here $n_k$ and $n$ depend
either on $t$ or $\tau$; that is, the scaling distribution $\Phi(x)$ is universal. Figure~\ref{Fig2} illustrates that for
the temperature-dependent agglomeration, \eqref{Sm1}--\eqref{energy},  the distribution $\Phi(x)$ quickly settles and
coincides with the one for the standard Smoluchowski equations.

The behavior of the scaled mass distribution $\Phi(x)$ depends on homogeneity indexes $\mu$ and $\nu$ defined via
\begin{equation}
\label{hom:mn}
K_{i,j}\sim i^\mu j^\nu  \quad\text{when}\quad j\gg i\,.
\end{equation}
Thus $\lambda=\mu+\nu$ and the reaction rates satisfying \eqref{hom} and \eqref{hom:mn} are   characterized by two
independent homogeneity indexes. For such homogeneous rates, qualitative behaviors are understood, see
\cite{Ernst1985,Ernst1988,connaughton2016universality,colmPRE2018}; it greatly depends on whether the index $\mu$ is
larger or smaller than zero. Reaction rates with $\mu<0$ are known as type III rates \cite{Ernst1985}.
The mass distribution in this case is bell-shaped \cite{Ernst1985,Ernst1988},
with an exponential decay in the large mass limit, and stretched exponential decay $\ln[1/\Phi(x)]\sim x^{-|\mu|}$ in the small mass limit.

For reaction rates \eqref{K:3} are of type III, the indexes are $\mu=-\frac{1}{2}$ and $\nu=2/3$.  Adopting the
treatment of Ref.~\cite{Ernst1988}  one can deduce a rather precise decay law
\begin{equation}
\label{n1:3d} -\ln [n_1(t)/n(0)] = C_1 t^{1/2}+C_2 t^{1/6} + O(1)
\end{equation}
for the density of monomers as we explain below.

\subsection{Ballistic aggregation: General dimension}

The generalization of Eqs.~\eqref{Sm1}--\eqref{energy} to arbitrary spatial dimension $d$ is straightforward. The
agglomeration rate is given by the same integral as in Eq.~\eqref{2} multiplied by $\Omega_{d-1}\sigma_{ij}^{d-1}$
instead of the factor $\pi \sigma_{ij}^2$ for the 3D systems. Here $\Omega_{d}=\pi^{d/2}/\Gamma(1+d/2)$ is the volume of a unit
$d$-dimensional ball.  Then one derives Eqs.~\eqref{Sm1} with mass-dependent rates
\begin{equation}
\label{K:d}
 K_{ij} = K_{0}(i^{1/d} + j^{1/d})^{d-1}   \sqrt{i^{-1} +j^{-1}}\,.
\end{equation}
Since the loss of energy in collisions is the same as in three dimensions, $\frac12\mu_{ij} v_{ij}^2$, the energy
equation  becomes
\begin{equation}
\label{energy-d} \frac{d}{dt}  nT = -\frac{d+1}{2d}\,T^{3/2} \sum_{i\geq 1} \sum_{j\geq 1}   K_{ij} n_in_j\,,
\end{equation}
where we have taken into account that $(d/2)nT$ gives the total kinetic energy in the $d$-dimensional case.

Using Eqs.~\eqref{Sm1}, \eqref{energy-d} and repeating analysis that has led to Eq.~\eqref{Tn-eq} we derive
\begin{equation}
\label{Tn-d-sol} \frac{T(t)}{T(0)} = \left[\frac{n(t)}{n(0)}\right]^{1/d}\,.
\end{equation}

The rates \eqref{K:d} are homogeneous, with homogeneity index $\lambda =(d-2)/(2d)$. The same analysis as in three
dimensions gives the asymptotic decay laws
\begin{equation}
\label{nT:d} n\sim t^{-2d/(d+3)}\,, \qquad \quad T\sim t^{-2/(d+3)}\,.
\end{equation}

The density of monomers in three dimensions decays according to Eq.~\eqref{n1:3d} in the large time limit. We now
derive this result, as well as the more general small mass asymptotic. We also outline a generalization to an arbitrary spatial
dimension. Our derivation adopts the procedure developed in Ref.~\cite{Ernst1988}. By inserting the scaling form \eqref{scaling} into
the Smoluchowski equations \eqref{Smol-eq} and using \eqref{K:d} we obtain
\begin{eqnarray}
\label{Smol-scaling}
w[2\Phi(x)+x\Phi'(x)] &=& \Phi(x)\int_0^\infty dy\,\Phi(y)\, K(x,y)\\
&-& \frac{1}{2}\int_0^x dy\,\Phi(y)\Phi(x-y)K(y,x-y)\nonumber
\end{eqnarray}
where
\begin{subequations}
\begin{align}
&\frac{dn}{d\tau} = - w n^{2-\lambda}\,, \qquad \lambda=\frac{1}{2}-\frac{1}{d}\\
&w = -\frac{1}{2}\int_0^\infty dx\int_0^\infty dy\,  \Phi(x)\, \Phi(y) \,K(x,y)\\
&K(x,y)  = K_{0}(x^{1/d} + y^{1/d})^{d-1}   \sqrt{x^{-1} +y^{-1}}
\end{align}
\end{subequations}
In the $y\to\infty$ limit, the kernel $K(x,y)$ admits an expansion
\begin{equation}
\label{Kxy} K(x,y)  = \sum_{n\geq 0} K_n x^{\mu_n} y^{\lambda-\mu_n}
\end{equation}
with $\mu_0=\mu=-\frac{1}{2}$ universal in all dimensions; $\mu_1=\frac{1}{d}-\frac{1}{2}$ and $K_1 = (d-1)K_0$;
$\mu_2=\frac{2}{d}-\frac{1}{2}, ~K_1 = \frac{1}{2}(d-1)(d-2)K_0$ when $d>2$ and $\mu_2=\frac{1}{2}, ~K_1 =
\frac{1}{2}K_0$ when $d=2$; etc. Inserting \eqref{Kxy} into \eqref{Smol-scaling} and focusing on the small mass
behavior, $x\downarrow 0$, we find
\begin{equation}
\label{Smol-small}
w[2\Phi(x)+x\Phi'(x)] \simeq \Phi(x) \sum_{n\geq 0} K_n x^{\mu_n} M_{\lambda-\mu_n}
\end{equation}
where $M_p$ is the $p^\text{th}$ moment of the scaled mass distribution:
\begin{equation}
M_p = \int_0^\infty dy\,\Phi(y)\, y^p
\end{equation}
Integrating Eq.~\eqref{Smol-small} one obtains \cite{Ernst1988}
\begin{equation}
\label{scaling-small}
\Phi(x)\sim x^{-2}\exp\!\left[\sum_{n\geq 0} \frac{K_n
M_{\lambda-\mu_n}}{w\mu_n}\,x^{\mu_n}\right]
\end{equation}
with sum running over such $n$ that $\mu_n < 0$ if $\mu_n\ne 0$ for all $n$; if $\mu_n = 0$ for some value $n$, the
term $x^{\mu_n}/\mu_n$ should be replaced by $\ln x$.

Since $\mu_0=-\frac{1}{2}, ~\mu_1=-\frac{1}{6}, \mu_2=\frac{1}{6}$ in three dimensions, Eq.~\eqref{scaling-small} becomes
\begin{equation}
\Phi(x)\sim x^{-2}\exp\!\left[-A_1x^{-1/2}-A_2x^{-1/6}\right].
\end{equation}
Thus $n_1=n^2\Phi(n)\sim \exp\!\left[A_1n^{-1/2}-A_2n^{-1/6}\right]$ leading to the announced asymptotic behavior \eqref{n1:3d} in three dimensions.

In two dimensions, $\mu_0=-\frac{1}{2}$ and $\mu_1=0$, so Eq.~\eqref{scaling-small} yields
$ n_1\sim \exp\!\left[-A_1n^{-1/2}-A_2\ln n\right]$, from which
\begin{equation}
\label{n1:2d}
-\ln\frac{n_1(t)}{n(0)} \sim C_1 t^\frac{2}{5} + C_2 \ln t + O(1)
\end{equation}
When $d=4$, we get $\mu_0=-\frac{1}{2}, ~\mu_1=-\frac{1}{4}$ and $\mu_2=0$, from which
\begin{equation}
\label{n1:4d}
-\ln\frac{n_1(t)}{n(0)} \sim C_1 t^\frac{4}{7} + C_2 t^\frac{2}{7} + C_3 \ln t + O(1)
\end{equation}
The constants $C_1, ~C_2$ etc. appearing in Eqs.~\eqref{n1:3d}, \eqref{n1:2d} are different, even if denoted by the same letter. These constants are unknown as they depend on the moments of the scaled mass distribution which are analytically unknown.

\begin{figure}
\centering
\includegraphics[width=8cm]{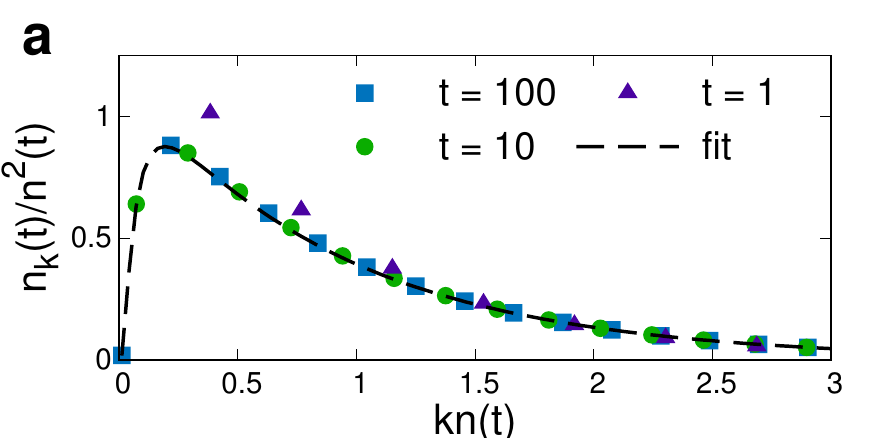}\\
\includegraphics[width=4cm]{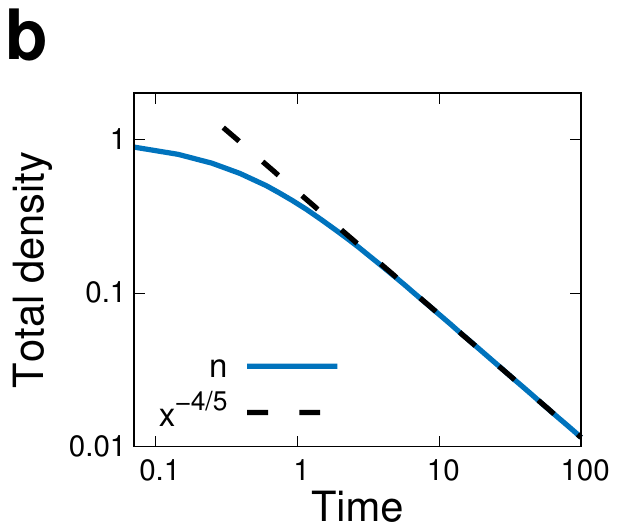}\q
\includegraphics[width=4cm]{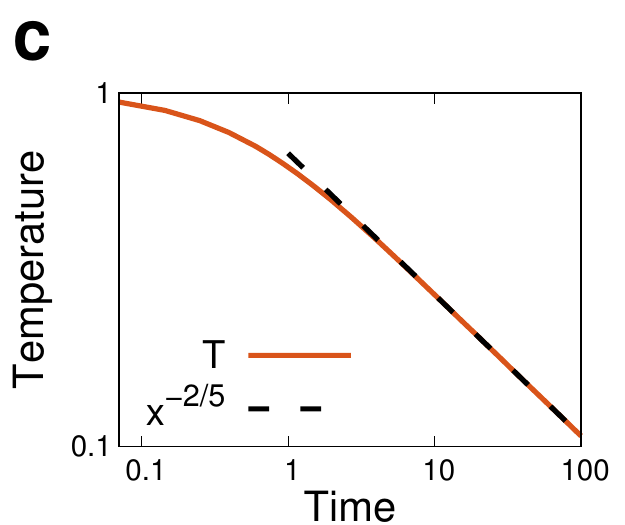}
\caption{(a) The reduced density distribution $n_k(t)/n^2(t)$ as a function of $kn(t)$ for different time  $t$ for 2D
system. The convergence to the scaling function $\Phi(x)$ with $x=kn(t)$ is observed already at $t = 10$. The scaled
mass distribution $\Phi(x)$ is well fit by $\left( \frac{2.13}{x^{0.25}} + \frac{1.03}{x^{1.28}} \right)
e^{-1.05/\sqrt{x}-1.035x}$. (b,c) The time dependence of the total density and temperature. Solid lines --- numerical results; dashed lines --- theoretical predictions,  Eqs.~\eqref{nT:d} for $d=2$. (a) and fit function are from the ODE solution; (b) and (c) are from Monte-Carlo simulations (see Appendix \ref{ap:num} for detail).}
 \label{SIFig1}
\end{figure}

Thus the monomer density exhibits the stretched exponential decay
\begin{equation}
\label{n1:d} \ln\frac{n_1(t)}{n(0)} \sim - t^\frac{d}{d+3}
\end{equation}
in the leading order. The leading behavior of the mass distribution in the small mass limit is
\begin{equation}
\label{nk:d} \ln\frac{n_k(t)}{n(0)} \sim - k^{-\frac{1}{2}}t^\frac{d}{d+3} \qquad\quad\text{for} \qquad \quad k\ll
t^\frac{2d}{d+3}\,.
\end{equation}
In Fig.~\ref{SIFig1} we present the density distribution and asymptotic behavior for $n(t)$ and $T(t)$ for
two-dimensional systems.

\section{Ballistic aggregation: Application to dark matter}
\label{sec:DM}

For many years, dark matter was thought of as a single stable and weakly interacting particle, but this paradigm is being
challenged by a wider view where dark matter is part of a  larger dark sector. In this framework, the formation of
dark nuclei with a very wide spectrum of masses becomes plausible. The agglomeration of dark nuclei from dark nucleons
has been studied in \cite{HEP1,HEP2,HEP3}. In our framework, the governing equations are \eqref{Sm1}--\eqref{energy},
with replacement $\frac{d}{dt}\to \frac{d}{dt} + 3H$, where $H=H(t)$ is the Hubble parameter accounting for the
expansion of the Universe. The transformation
\begin{equation}
\label{nhm}
 n_k = h m_k, \qquad h(t) =\exp\!\left[-3\int_{t_0}^t dt'\,H(t')\right]
\end{equation}
recasts these equations into
\begin{eqnarray}
\label{Sm-h} \frac{1}{h  T^{1/2}}\frac{d m_k}{dt}\!\! &= &\!\!\frac{1}{2} \!\!\sum_{i+j=k} \!\! K_{ij} m_i m_j \!-\!
m_k \!\sum_{i\geq 1}\!K_{ki} m_i\\
\label{energy-h} \frac{1}{h  T^{3/2}}\frac{d\,mT}{dt} \!\!& = &\! \! - \frac23\sum_{i\geq 1} \sum_{j\geq 1}   K_{ij}
m_i m_j
\end{eqnarray}
that differ from \eqref{Sm1}--\eqref{energy} only by an extra factor $h(t)$. In Ref.~\cite{HEP1} it was assumed that
the dark nuclei were in contact with a bath of lighter particles, which determined their temperature. The temperature of the bath was gradually decreasing during the evolution of the Universe. Here we only take into account collisions between dark nuclei, so the temperature is defined by the agglomeration and Hubble expansion only;  that is the system of dark nucleons is assumed to be completely isolated.

Agglomeration begins at sufficiently low temperatures, say when the temperature drops below $T_0$. Initially, the temperature decreases mainly due to radiation, which is especially important at high temperatures in the early stages of the Universe. However, we assume that at $T=T_0$ this type of energy loss is already quite slow, so that the aggregation quickly becomes dominant when it starts. In the definition
\eqref{nhm} of $h(t)$ we set the lower limit $t_0$ as the time when this occurs,  $T_0=T(t_0)$. The natural initial
condition is $m_k(t_0)=n_0\delta_{k,1}$, where  $n_0 = n(t_0)$. Using \eqref{Sm-h}--\eqref{energy-h} we find that for
$t\geq t_0$ the temperature and the auxiliary total density are related via
\begin{equation}
\label{Tm-sol}
T(t)/T_0= \left[m(t)/n_0\right]^{1/3}.
\end{equation}
We rescale  $m_k \to n_0 m_k$, $T \to T_0 T$ and $K_0 \to n_0T_0^{1/2}K_0$, where $K_0$ is defined by Eq. \eqref{K:3}
and keep, for simplicity,  the same notations for these quantities. Then with the dimensionless time
\begin{equation}
\label{tau:h} {\cal T} = K_0 \, \int_{t_0}^t dt'\, h(t')
\end{equation}
we recast Eqs.~\eqref{Sm-h}--\eqref{energy-h} into the temperature-dependent Smoluchowski equations
\eqref{Sm1}--\eqref{energy} for $m_k({\cal T} )$, whose properties have been analyzed previously.

To determine $h(t)$, we need a bit of cosmology.  There is solid observational evidence in favor of the
flat Universe with positive cosmological constant $\Lambda$ representing dark energy. Then the Friedmann equation for
the scaled factor $a(t)$ reads
\begin{equation}
\label{Friedmann-DE}
\frac{\dot a^2}{a^2} = \frac{8\pi G \rho + \Lambda c^2}{3}\,.
\end{equation}
Here $G$ is the Newton constant, $\rho$ the density, $c$ the speed of light and $H=a^{-1}\dot a$.
Density $\rho$ can be determined from the Friedman acceleration equation
\begin{equation}
\label{Friedmann-2}
\frac{\ddot a}{a} = - \frac{4 \pi G}{3} \left(  \rho + \frac{3p}{c^2} \right) + \frac{\Lambda c^2}{3},
\end{equation}
where $p$ is pressure. Combining \eqref{Friedmann-2} with \eqref{Friedmann-DE} we find $\dot \rho = -\frac{3 \dot a}{a} \left(
\rho + p/ c^2 \right)$. If agglomeration of dark matter indeed occurs, it begins in the radiation dominated era of the
expansion \cite{HEP1}. At this stage the equation of state is $p=\rho c^2/3$ and from the previous equation for
$\dot{\rho}$ one finds $\rho(t)/\rho_0 = \left(a_0/a(t) \right)^{4}$. Using this result together  with $\Lambda c^2 \ll 8 \pi
G \rho$ which is valid in the radiation era, see \cite{NOHEP}, we simplify Eq.~\eqref{Friedmann-DE} to
\begin{equation}
\label{Friedmann}
a\,\frac{da}{dt} = \sqrt{\frac{8\pi G \rho_0 a_0^4}{3}}
\end{equation}
Integrating \eqref{Friedmann}, using Eq. \eqref{nhm} with  $H=a^{-1}\dot a$ we find
\begin{equation}
\label{h-sol-DE} h(t) = \left[1 + 2 H_0(t-t_0) \right]^{-3/2}
\end{equation}
with $H_0 =H(t_0) = \left(8 \pi G \rho_0/ 3\right)^{1/2}$. Equation \eqref{tau:h} then yields
\begin{equation}
\label{tau-t} {\cal T} (t) = \frac{K_0}{H_0} \left( 1 - \frac{1}{\sqrt{1+2H_0(t-t_0) }} \right).
\end{equation}
The modified time ${\cal T} $ remains finite and the agglomeration effectively ceases in the
radiation-dominated era if $H_0(t_r-t_0)\gg 1$.

\begin{figure}
\centering
\includegraphics[width=8.5cm]{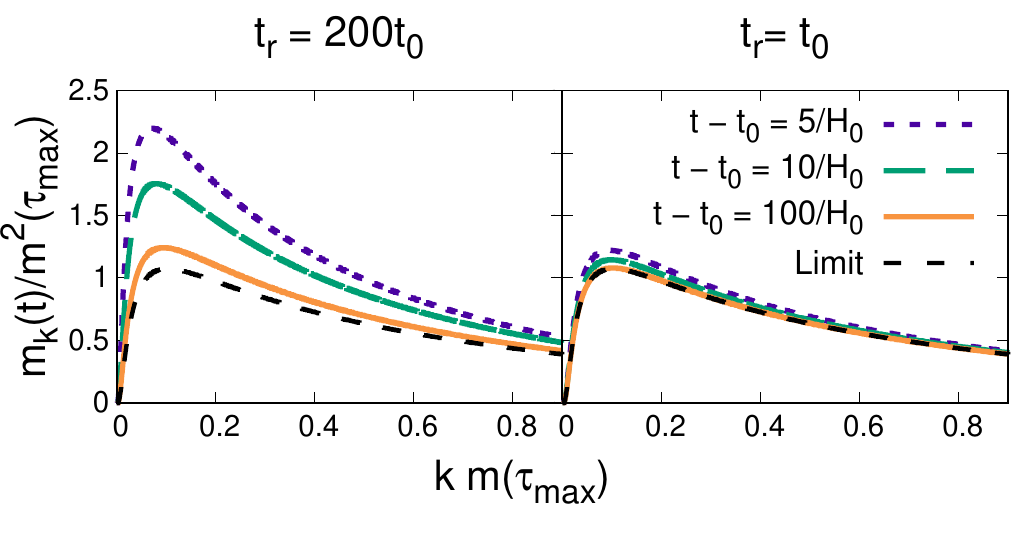}
\caption{The modified densities $m_k(t)$ as a function  of size $k$ for different agglomeration time $t-t_0$. For
$t_r=200 \, t_0$ the agglomeration ceases in the radiation-dominated era (left), while for $t_0=t_r$ -- in the
matter-dominated era (right). In both cases the distribution converges to the frozen one \eqref{mkfroz}. The particular
value $m({\cal T}_{\rm max})$ is determined by $K_0/H \simeq 2K_0t_0$, which is assumed to be large enough to guarantee
scaling. Results are obtained by the ODE solution (see Appendix \ref{ap:num}).}
 \label{fig3}
\end{figure}

If $t_0$ is close to the end of the radiation era $t_r$,  the agglomeration
continues for $t>t_r$. The freezing then occurs in the matter-dominated era, so one can use \cite{NOHEP}
\begin{equation}
a^{1/2}\,\frac{da}{dt} = \sqrt{\frac{8\pi G \rho_0 a_0^3}{3}}.
\end{equation}
Solving this equation one can find $H(t) = \dot{a}(t)/a(t)$ for the matter-dominated era, $h(t)$ and eventually ${\cal
T}(t)$. The details of the derivation are presented in Appendix \ref{ap:dark}, here we just quote the result:
\begin{equation}
\label{modtlast} {\cal T}(t) = 2K_0t_0 \left[ 1- \sqrt{\frac{t_0}{t_r}} \left(\frac{3t_r+t}{3t+t_r} \right) \right] \,.
\end{equation}
For simplicity, we have assumed a sharp transition from the radiation to matter-dominated era. Hence the modified time
remains finite:
\begin{equation}
\label{tautrmax} {\cal T} < {\cal T}_{\max} = 2K_0t_0 \left(1 - \sqrt{t_0/9t_r} \right).
 \end{equation}
The evolution thus {\em freezes}, and if ${\cal T}_{\max}$ is large enough, the modified densities, $m_k({\cal T})$,
the frozen scaled form
\begin{equation} \label{mkfroz}
 m_k^{\rm frozen} = m^2 ({\cal T}_{\rm max})
\Phi \left[ k m({\cal T}_{\rm max}) \right],
\end{equation}
where the scaling function $\Phi(x)$ is the same as for the standard or temperature-dependent Smoluchowski equations.
Evolution with the freezing has been also reported in \cite{HEP1}. Figure \ref{fig3} illustrates the convergence of $m_k$
to the frozen distribution \eqref{mkfroz} for increasing $t$ for two scenarios: the freezing within the
radiation-dominated  and matter-dominates eras.

\section{Conclusion}
\label{sec:concl}

We have investigated the ballistic agglomeration process in the reaction-controlled limit. Cluster densities satisfy an infinite set of
Smoluchowski rate equations, with rates proportional $\sqrt{T}$, where $T$ is the kinetic temperature whose evolution
is described by an energy equation.  Remarkably, the temperature admits an expression through the total cluster density
alone. In the reaction-controlled limit, the exponents describing the evolution of the total density and energy have
been established in Ref.~\cite{Trizac:2003}. Our more comprehensive description additionally gives the mass
distribution. In particular, we have obtained an unexpected stretched exponential decay for the density of monomers.  Our
theoretical findings are in good agreement with simulation results.

We emphasize that the ballistic agglomeration process in the collision-controlled limit is not yet analytically
understood in three dimensions, and generally when $d\geq 2$;  the one-dimensional model is exactly solvable
\cite{Frachebourg1999,FrachebourgPiasecki2000,LF00,BA1D}. Some quantities exhibit drastically different behaviors in
the reaction-controlled and collision-controlled cases. For instance, in one dimension the density of monomers decays
as $\exp\!\left[-Ct^{1/4}\right]$ in the reaction-controlled case, while in the collision-controlled limit  $n_1\sim
t^{-1}$, that is the decay is much slower.

We have applied our formalism to the evolution of dark matter, namely to  a model of asymmetric dark matter
\cite{HEP1,HEP2,HEP3} where dark nuclei are formed via agglomeration of elementary dark nucleons. We have assumed that
collision events rarely lead to merging. In this reaction-controlled limit, the system reaches the temperature equipartition for different cluster species without the need for the bath of light particles \cite{HEP1,HEP3}. The size distribution of the dark nuclei tends to a
final frozen distribution whose functional form follows from the solution of the standard Smoluchowski equations. A
wide spectrum of masses calls for novel strategies for direct detection of heavy dark matter nuclei \cite{Zurek19}.
Among possible directions for future work, we mention symmetric dark matter models where the agglomeration should be supplemented
by annihilation.

\appendix
{
\section{Numerical methods} 
\label{ap:num}

In our study we used two different numerical methods: the ODE solution and Monte-Carlo simulations. Both methods are popular and efficient tools to study the aggregation kinetics, see e.g.  \cite{Family1985,ThornPRL1994,Odriozola:2001,Odriozola:2004}.

Monte-Carlo simulations have been used to obtain the results for Fig. \ref{Fig2}(b) and Fig. \ref{SIFig1}(b,c); it allows directly prove the validity of Eq.~\eqref{Tn-d-sol}. The detail of the Monte-Carlo approach exploited here may be found in Refs.~\cite{bodrova2020, bodrovamol}. The only difference of the present implementation of this method is that the speeds of the particles were generated from the Maxwell distribution before each collision, without the use of a bath, as in  \cite{bodrova2020, bodrovamol}. This follows from the fact that the particles have time to exchange kinetic energy between the aggregation events. We used $10^7$ particles and doubled them every time when their number decreased by a factor of two.

In other figures (Fig. \ref{Fig2}(a,c), Fig. \ref{SIFig1}(a), Fig. \ref{fig3}) we solved the ODE system \eqref{Sm1}--\eqref{energy} directly, after limiting the total number of equations to 50000. While Monte-Carlo and ODE method converge to the same solution, when the time step goes to zero and the number of particles goes to infinity, the ODE solution converges much faster and does not have stochastic noise. For better accuracy, we used second-order predictor-corrector scheme with an adaptive time step of $\tau = 0.01$. In this case, the time step is  calculated as
\begin{equation}
\begin{aligned}
  \Delta t_k & = \tau \frac{\max\limits_i \left| n_i(t_{k-1}) \right|}{\max\limits_i \left| n_i(t_{k-1}) - n_i(t_k) \right|}, \\
  t_{k+1} & = t_k + \Delta t_k.
\end{aligned}
\end{equation}
To speed up the solution of the ODE system, we used the method for generalized Smoluchowski equations \cite{gen-smol}, which is based on a low-rank approach from \cite{matveev,matveev1}. The same solution can be obtained by solving the Smoluchowski equations directly like any other finite system of ODEs. However, the application of the low-rank approximation and adaptive time step technique accelerates the computations enormously \cite{gen-smol, matveev,matveev1}.
}

\section{Derivation of Eq.~(43) }
\label{ap:dark}

If the agglomeration starting time  $t_0$ is close to the end of the radiation-dominated era, $t_r \approx 50000 \text{ years}$, a significant number of collisions still happen for $t>t_r$ in the matter-dominated era. For $t>t_r$, the pressure becomes very small, so $\dot \rho = -\frac{3 \dot a}{a}  \left( \rho + p/ c^2 \right)$ reduces to  $\dot \rho = -3 (\dot a/a) \rho$.

Obviously, the transition from the state with the non-vanishing pressure $p_r$ in the radiation-dominated era to the
state with $p \approx 0$  in the matter-dominated era is not instantaneous. The evolution of pressure for the transient
period may be described (see \cite{NOHEP}) as
$$
p \approx \frac{p_{r}}{1 + T_{eq}/T}\,,
$$
where $T_{eq}$ is the temperature of matter-radiation equilibrium. This makes the analysis of the transient period
extremely complicated and does not allow us to obtain an explicit relation for the modified time $\tau$. Therefore, for
the qualitative analysis, we assume that the transition period is short enough, as compared to the total time of the
formation of the density distribution of dark matter. Hence we effectively postulate that this transition is instantaneous. Combining
$\rho(t)/\rho(t_r) = \left( a(t_r) / a(t) \right)^{3}$ and Eq.~\eqref{Friedmann-DE} we obtain
\begin{equation}
\label{Friedmann2} a^{1/2}\,\frac{da}{dt} = \sqrt{\frac{8\pi G \rho_r a_r^3}{3}}
\end{equation}
for $t > t_r$,  where $\rho_r$ and $a_r$ are the quantities at the end of the radiation-dominated era. This equation is solved to yield the behavior $H(t) =\dot{a}(t)/a(t)$ for the matter-dominated era. The quantity $h(t)$ defined in Eq.~\eqref{nhm} reads
\begin{equation}\label{ht2}
  h(t) = h_r\left[1 + \frac{3}{2} H_r \left( t - t_r \right) \right]^{-2}, \qquad t > t_r,
\end{equation}
where $H_r=H(t_r)$ and $h_r=h(t_r)$ are again the quantities at the end of the radiation-dominated era.
Note that one can ignore the cosmological constant $\Lambda$ which becomes relevant only when the Universe is older than about 10 billion years.

To determine the modified time ${\cal T}$ for $t>t_r$ we first need to find ${\cal T}(t_r)$ from Eq.~\eqref{Friedmann2}
and then add the corresponding integral with  $h(t)$ given by Eq.~\eqref{ht2}. To simplify the computations, let us assume
that $t_0$ is far from cosmic inflation (otherwise, everything would have already aggregated in radiation-dominated
era). Then since $a(t) \sim t^{1/2}$, as it follows from Eq.~\eqref{Friedmann} we obtain the estimates for the
beginning of the agglomeration and the end of the radiation era:
\begin{eqnarray}
  H_0 = \frac{\dot a(t_0)}{a_0}  \simeq \frac{1}{2t_0}\,, \qquad
H_r \simeq  \frac{1}{2t_r }
\end{eqnarray}
Then we arrive at
\begin{eqnarray*}
  {\cal T}(t) & =& {\cal T}(t_r) +  \left[ {\cal T}(t) -{\cal T}(t_r) \right]  \\
  & =& \frac{K_0}{H_0} \left(1 - \frac{1}{\sqrt{1 + 2 H_0 \left(t_r - t_0 \right)}} \right) +
  \int_{t_r}^t dt'\,{K_0} h(t')
\end{eqnarray*}
Massaging this expression we simplify it to
\begin{eqnarray}
  {\cal T}(t)   & \simeq & 2K_0t_0 \left( 1 - \sqrt{\frac{t_0}{t_r}} \right) \nonumber \\
  &+& \frac{4{K_0}t_r}{3} \left( \frac{t_0}{t_r} \right)^{3/2}
  \left( 1 - \frac{1}{1 + \frac{3}{4t_r} \left(t - t_r \right)} \right) \nonumber \\
    &=& 2K_0t_0 \left[ 1- \sqrt{\frac{t_0}{t_r}} \left(\frac{3t_r+t}{3t+t_r} \right) \right] . \nonumber
\end{eqnarray}
This completes the derivation of Eq.~\eqref{modtlast}. In the long time limit, one can further simplify to obtain Eq.~\eqref{tautrmax}.

\vskip 1cm
\noindent 
{\bf Acknowledgments.}  This study has been supported by RFBR through the research projects  \textnumero 18-29-1919 and   \textnumero 20-31-90022.

\bibliography{agglomeration}

\begin{thebibliography}{54}
\expandafter\ifx\csname natexlab\endcsname\relax\def\natexlab#1{#1}\fi
\expandafter\ifx\csname bibnamefont\endcsname\relax
  \def\bibnamefont#1{#1}\fi
\expandafter\ifx\csname bibfnamefont\endcsname\relax
  \def\bibfnamefont#1{#1}\fi
\expandafter\ifx\csname citenamefont\endcsname\relax
  \def\citenamefont#1{#1}\fi
\expandafter\ifx\csname url\endcsname\relax
  \def\url#1{\texttt{#1}}\fi
\expandafter\ifx\csname urlprefix\endcsname\relax\def\urlprefix{URL }\fi
\providecommand{\bibinfo}[2]{#2}
\providecommand{\eprint}[2][]{\url{#2}}

\bibitem[{\citenamefont{Smoluchowski}(1917)}]{Smo17}
\bibinfo{author}{\bibfnamefont{M.~V.} \bibnamefont{Smoluchowski}},
  \bibinfo{journal}{Z. Phys. Chem.} \textbf{\bibinfo{volume}{92}},
  \bibinfo{pages}{129} (\bibinfo{year}{1917}).

\bibitem[{\citenamefont{Family et~al.}(1985)\citenamefont{Family, Meakin, and
  Vicsek}}]{Family1985}
\bibinfo{author}{\bibfnamefont{F.}~\bibnamefont{Family}},
  \bibinfo{author}{\bibfnamefont{P.}~\bibnamefont{Meakin}}, \bibnamefont{and}
  \bibinfo{author}{\bibfnamefont{T.}~\bibnamefont{Vicsek}},
  \bibinfo{journal}{J. Chem. Phys.} \textbf{\bibinfo{volume}{83}},
  \bibinfo{pages}{4144} (\bibinfo{year}{1985}).

\bibitem[{\citenamefont{Ball et~al.}(1987)\citenamefont{Ball, Weitz, Witten,
  and Leyvraz}}]{Ball1987}
\bibinfo{author}{\bibfnamefont{R.~C.} \bibnamefont{Ball}},
  \bibinfo{author}{\bibfnamefont{D.~A.} \bibnamefont{Weitz}},
  \bibinfo{author}{\bibfnamefont{T.~A.} \bibnamefont{Witten}},
  \bibnamefont{and} \bibinfo{author}{\bibfnamefont{F.}~\bibnamefont{Leyvraz}},
  \bibinfo{journal}{Phys. Rev. Lett.} \textbf{\bibinfo{volume}{58}},
  \bibinfo{pages}{274} (\bibinfo{year}{1987}).

\bibitem[{\citenamefont{Thorn and Seesselberg}(1994)}]{ThornPRL1994}
\bibinfo{author}{\bibfnamefont{M.}~\bibnamefont{Thorn}} \bibnamefont{and}
  \bibinfo{author}{\bibfnamefont{M.}~\bibnamefont{Seesselberg}},
  \bibinfo{journal}{Phys. Rev. Lett.} \textbf{\bibinfo{volume}{72}},
  \bibinfo{pages}{3622} (\bibinfo{year}{1994}).

\bibitem[{\citenamefont{Odriozola et~al.}(2001)\citenamefont{Odriozola,
  Moncho-Jord\'{a}, Schmitt, Callejas-Fern\'{a}ndez, Mart\'{i}nez-Garc\'{i}a,
  and Hidalgo-\'{A}lvarez}}]{Odriozola:2001}
\bibinfo{author}{\bibfnamefont{G.}~\bibnamefont{Odriozola}},
  \bibinfo{author}{\bibfnamefont{A.}~\bibnamefont{Moncho-Jord\'{a}}},
  \bibinfo{author}{\bibfnamefont{A.}~\bibnamefont{Schmitt}},
  \bibinfo{author}{\bibfnamefont{J.}~\bibnamefont{Callejas-Fern\'{a}ndez}},
  \bibinfo{author}{\bibfnamefont{R.}~\bibnamefont{Mart\'{i}nez-Garc\'{i}a}},
  \bibnamefont{and}
  \bibinfo{author}{\bibfnamefont{R.}~\bibnamefont{Hidalgo-\'{A}lvarez}},
  \bibinfo{journal}{Europhys. Lett.} \textbf{\bibinfo{volume}{53}},
  \bibinfo{pages}{797} (\bibinfo{year}{2001}).

\bibitem[{\citenamefont{Odriozola et~al.}(2004)\citenamefont{Odriozola, Leone,
  Schmitt, Callejas-Fern\'{a}ndez, Mart\'{i}nez-Garc\'{i}a, and
  Hidalgo-\'{A}lvarez}}]{Odriozola:2004}
\bibinfo{author}{\bibfnamefont{G.}~\bibnamefont{Odriozola}},
  \bibinfo{author}{\bibfnamefont{R.}~\bibnamefont{Leone}},
  \bibinfo{author}{\bibfnamefont{A.}~\bibnamefont{Schmitt}},
  \bibinfo{author}{\bibfnamefont{J.}~\bibnamefont{Callejas-Fern\'{a}ndez}},
  \bibinfo{author}{\bibfnamefont{R.}~\bibnamefont{Mart\'{i}nez-Garc\'{i}a}},
  \bibnamefont{and}
  \bibinfo{author}{\bibfnamefont{R.}~\bibnamefont{Hidalgo-\'{A}lvarez}},
  \bibinfo{journal}{J. Chem. Phys.} \textbf{\bibinfo{volume}{121}},
  \bibinfo{pages}{5468} (\bibinfo{year}{2004}).

\bibitem[{\citenamefont{Flory}(1953)}]{Flory}
\bibinfo{author}{\bibfnamefont{P.~J.} \bibnamefont{Flory}},
  \emph{\bibinfo{title}{Principles of Polymer Chemistry}}
  (\bibinfo{publisher}{Cornell University Press}, \bibinfo{year}{1953}).

\bibitem[{\citenamefont{Hidy and Brock}(1970)}]{Hidy}
\bibinfo{author}{\bibfnamefont{G.~R.} \bibnamefont{Hidy}} \bibnamefont{and}
  \bibinfo{author}{\bibfnamefont{J.~R.} \bibnamefont{Brock}},
  \emph{\bibinfo{title}{The Dynamics of Aerocolloidal Systems, International
  Reviews in Aerosol Physics and Chemistry}} (\bibinfo{publisher}{Pergamon
  Press}, \bibinfo{address}{Oxford}, \bibinfo{year}{1970}).

\bibitem[{\citenamefont{Drake}(1972)}]{Drake}
\bibinfo{author}{\bibfnamefont{R.~L.} \bibnamefont{Drake}},
  \emph{\bibinfo{title}{In: G. M. Hidy, and J. R. Brock (Eds.), Topics in
  Current Aerosol Research, Vol. 3, part 2}} (\bibinfo{publisher}{Pergamon
  Press}, \bibinfo{address}{New York}, \bibinfo{year}{1972}).

\bibitem[{\citenamefont{Shrivastava}(1982)}]{Srivastava1982}
\bibinfo{author}{\bibfnamefont{R.~C.} \bibnamefont{Shrivastava}},
  \bibinfo{journal}{J. Atom. Sci.} \textbf{\bibinfo{volume}{39}},
  \bibinfo{pages}{1317} (\bibinfo{year}{1982}).

\bibitem[{\citenamefont{Friedlander}(2000)}]{Friedlander}
\bibinfo{author}{\bibfnamefont{S.~K.} \bibnamefont{Friedlander}},
  \emph{\bibinfo{title}{Smoke, Dust and Haze, 2nd Edition}}
  (\bibinfo{publisher}{Oxford University Press}, \bibinfo{address}{Oxford},
  \bibinfo{year}{2000}).

\bibitem[{\citenamefont{Field and Saslaw}(1965)}]{Saslaw}
\bibinfo{author}{\bibfnamefont{G.~B.} \bibnamefont{Field}} \bibnamefont{and}
  \bibinfo{author}{\bibfnamefont{W.~C.} \bibnamefont{Saslaw}},
  \bibinfo{journal}{Astrophys. J} \textbf{\bibinfo{volume}{142}},
  \bibinfo{pages}{568} (\bibinfo{year}{1965}).

\bibitem[{\citenamefont{Lissauer}(1993)}]{Lissauer:1993}
\bibinfo{author}{\bibfnamefont{J.~J.} \bibnamefont{Lissauer}},
  \bibinfo{journal}{Ann. Rev. Astron. Astrophys.}
  \textbf{\bibinfo{volume}{31}}, \bibinfo{pages}{129} (\bibinfo{year}{1993}).

\bibitem[{\citenamefont{Chokshi et~al.}(1993)\citenamefont{Chokshi, Tielens,
  and Hollenbach}}]{Chokshietal:1993}
\bibinfo{author}{\bibfnamefont{A.}~\bibnamefont{Chokshi}},
  \bibinfo{author}{\bibfnamefont{A.~G.~G.} \bibnamefont{Tielens}},
  \bibnamefont{and}
  \bibinfo{author}{\bibfnamefont{D.}~\bibnamefont{Hollenbach}},
  \bibinfo{journal}{Astrophys. J.} \textbf{\bibinfo{volume}{407}},
  \bibinfo{pages}{806} (\bibinfo{year}{1993}).

\bibitem[{\citenamefont{Dominik and Tielens}(1997)}]{DominikTilens:1997}
\bibinfo{author}{\bibfnamefont{C.}~\bibnamefont{Dominik}} \bibnamefont{and}
  \bibinfo{author}{\bibfnamefont{A.~G.~G.} \bibnamefont{Tielens}},
  \bibinfo{journal}{Astrophys. J.} \textbf{\bibinfo{volume}{480}},
  \bibinfo{pages}{647} (\bibinfo{year}{1997}).

\bibitem[{\citenamefont{Ossenkopf}(1993)}]{Ossenkopf1993}
\bibinfo{author}{\bibfnamefont{V.}~\bibnamefont{Ossenkopf}},
  \bibinfo{journal}{Astron. Astrophys.} \textbf{\bibinfo{volume}{280}}
  (\bibinfo{year}{1993}).

\bibitem[{\citenamefont{Spahn et~al.}(2004)\citenamefont{Spahn, Albers,
  Sremcevic, and Thornton}}]{SpahnAlbersetal:2004}
\bibinfo{author}{\bibfnamefont{F.}~\bibnamefont{Spahn}},
  \bibinfo{author}{\bibfnamefont{N.}~\bibnamefont{Albers}},
  \bibinfo{author}{\bibfnamefont{M.}~\bibnamefont{Sremcevic}},
  \bibnamefont{and} \bibinfo{author}{\bibfnamefont{C.}~\bibnamefont{Thornton}},
  \bibinfo{journal}{Europhys. Lett.} \textbf{\bibinfo{volume}{67}},
  \bibinfo{pages}{545} (\bibinfo{year}{2004}).

\bibitem[{\citenamefont{Esposito}(2006)}]{esposito2006}
\bibinfo{author}{\bibfnamefont{L.}~\bibnamefont{Esposito}},
  \emph{\bibinfo{title}{Planetary Rings}} (\bibinfo{publisher}{Cambridge
  University Press}, \bibinfo{address}{Cambridge, UK}, \bibinfo{year}{2006}).

\bibitem[{\citenamefont{Brilliantov et~al.}(2015)\citenamefont{Brilliantov,
  Krapivsky, Bodrova, Spahn, Hayakawa, Stadnichuk, and
  Schmidt}}]{BrilliantovPNAS2015}
\bibinfo{author}{\bibfnamefont{N.~V.} \bibnamefont{Brilliantov}},
  \bibinfo{author}{\bibfnamefont{P.~L.} \bibnamefont{Krapivsky}},
  \bibinfo{author}{\bibfnamefont{A.}~\bibnamefont{Bodrova}},
  \bibinfo{author}{\bibfnamefont{F.}~\bibnamefont{Spahn}},
  \bibinfo{author}{\bibfnamefont{H.}~\bibnamefont{Hayakawa}},
  \bibinfo{author}{\bibfnamefont{V.}~\bibnamefont{Stadnichuk}},
  \bibnamefont{and} \bibinfo{author}{\bibfnamefont{J.}~\bibnamefont{Schmidt}},
  \bibinfo{journal}{Proc. Natl. Acad. Sci. USA} \textbf{\bibinfo{volume}{112}},
  \bibinfo{pages}{9536} (\bibinfo{year}{2015}).

\bibitem[{\citenamefont{Hardy et~al.}(2015)\citenamefont{Hardy, Lasenby,
  March-Russell, and West}}]{HEP1}
\bibinfo{author}{\bibfnamefont{E.}~\bibnamefont{Hardy}},
  \bibinfo{author}{\bibfnamefont{R.}~\bibnamefont{Lasenby}},
  \bibinfo{author}{\bibfnamefont{J.}~\bibnamefont{March-Russell}},
  \bibnamefont{and} \bibinfo{author}{\bibfnamefont{S.~W.} \bibnamefont{West}},
  \bibinfo{journal}{JHEP} \textbf{\bibinfo{volume}{06}}, \bibinfo{pages}{011}
  (\bibinfo{year}{2015}).

\bibitem[{\citenamefont{Krnjaic and Sigurdson}(2015)}]{HEP2}
\bibinfo{author}{\bibfnamefont{G.}~\bibnamefont{Krnjaic}} \bibnamefont{and}
  \bibinfo{author}{\bibfnamefont{K.}~\bibnamefont{Sigurdson}},
  \bibinfo{journal}{Physics Letters B} \textbf{\bibinfo{volume}{751}},
  \bibinfo{pages}{464} (\bibinfo{year}{2015}).

\bibitem[{\citenamefont{Gresham et~al.}(2018)\citenamefont{Gresham, Lou, and
  Zurek}}]{HEP3}
\bibinfo{author}{\bibfnamefont{M.~I.} \bibnamefont{Gresham}},
  \bibinfo{author}{\bibfnamefont{H.~K.} \bibnamefont{Lou}}, \bibnamefont{and}
  \bibinfo{author}{\bibfnamefont{K.~M.} \bibnamefont{Zurek}},
  \bibinfo{journal}{Phys. Rev. D} \textbf{\bibinfo{volume}{97}},
  \bibinfo{pages}{036003} (\bibinfo{year}{2018}).

\bibitem[{\citenamefont{Leyvraz}(2003)}]{Leyvraz2003}
\bibinfo{author}{\bibfnamefont{F.}~\bibnamefont{Leyvraz}},
  \bibinfo{journal}{Physics Reports} \textbf{\bibinfo{volume}{383}},
  \bibinfo{pages}{95} (\bibinfo{year}{2003}).

\bibitem[{\citenamefont{Krapivsky et~al.}(2010)\citenamefont{Krapivsky, Redner,
  and Ben-Naim}}]{krapbook}
\bibinfo{author}{\bibfnamefont{P.~L.} \bibnamefont{Krapivsky}},
  \bibinfo{author}{\bibfnamefont{A.}~\bibnamefont{Redner}}, \bibnamefont{and}
  \bibinfo{author}{\bibfnamefont{E.}~\bibnamefont{Ben-Naim}},
  \emph{\bibinfo{title}{A Kinetic View of Statistical Physics}}
  (\bibinfo{publisher}{Cambridge University Press},
  \bibinfo{address}{Cambridge, UK}, \bibinfo{year}{2010}).

\bibitem[{\citenamefont{Spouge}(1983)}]{bilinearkernel2}
\bibinfo{author}{\bibfnamefont{J.}~\bibnamefont{Spouge}}, \bibinfo{journal}{J.
  Phys. A} \textbf{\bibinfo{volume}{16}}, \bibinfo{pages}{3127}
  (\bibinfo{year}{1983}).

\bibitem[{\citenamefont{Calogero and Leyvraz}(2000)}]{paritykernel}
\bibinfo{author}{\bibfnamefont{F.}~\bibnamefont{Calogero}} \bibnamefont{and}
  \bibinfo{author}{\bibfnamefont{F.}~\bibnamefont{Leyvraz}},
  \bibinfo{journal}{J. Phys. A} \textbf{\bibinfo{volume}{33}},
  \bibinfo{pages}{5619} (\bibinfo{year}{2000}).

\bibitem[{\citenamefont{Calogero and Leyvraz}(1999)}]{qsumkernel}
\bibinfo{author}{\bibfnamefont{F.}~\bibnamefont{Calogero}} \bibnamefont{and}
  \bibinfo{author}{\bibfnamefont{F.}~\bibnamefont{Leyvraz}},
  \bibinfo{journal}{J. Phys. A} \textbf{\bibinfo{volume}{32}},
  \bibinfo{pages}{7697} (\bibinfo{year}{1999}).

\bibitem[{\citenamefont{van Dongen and Ernst}(1985)}]{Ernst1985}
\bibinfo{author}{\bibfnamefont{P.~G.~J.} \bibnamefont{van Dongen}}
  \bibnamefont{and} \bibinfo{author}{\bibfnamefont{M.~H.} \bibnamefont{Ernst}},
  \bibinfo{journal}{Phys. Rev. Lett.} \textbf{\bibinfo{volume}{54}},
  \bibinfo{pages}{1396} (\bibinfo{year}{1985}).

\bibitem[{\citenamefont{van Dongen and Ernst}(1988)}]{Ernst1988}
\bibinfo{author}{\bibfnamefont{P.~G.~J.} \bibnamefont{van Dongen}}
  \bibnamefont{and} \bibinfo{author}{\bibfnamefont{M.~H.} \bibnamefont{Ernst}},
  \bibinfo{journal}{J. Stat. Phys.} \textbf{\bibinfo{volume}{50}},
  \bibinfo{pages}{295} (\bibinfo{year}{1988}).

\bibitem[{\citenamefont{Carnevale et~al.}(1990)\citenamefont{Carnevale, Pomeau,
  and Young}}]{CarnevalePameauYoung:1990}
\bibinfo{author}{\bibfnamefont{G.~F.} \bibnamefont{Carnevale}},
  \bibinfo{author}{\bibfnamefont{Y.}~\bibnamefont{Pomeau}}, \bibnamefont{and}
  \bibinfo{author}{\bibfnamefont{W.~R.} \bibnamefont{Young}},
  \bibinfo{journal}{Phys. Rev. Lett.} \textbf{\bibinfo{volume}{64}},
  \bibinfo{pages}{2913} (\bibinfo{year}{1990}).

\bibitem[{\citenamefont{Trizac and Hansen}(1995)}]{TrizacHansen:1995}
\bibinfo{author}{\bibfnamefont{E.}~\bibnamefont{Trizac}} \bibnamefont{and}
  \bibinfo{author}{\bibfnamefont{J.-P.} \bibnamefont{Hansen}},
  \bibinfo{journal}{Phys. Rev. Lett.} \textbf{\bibinfo{volume}{74}},
  \bibinfo{pages}{4114} (\bibinfo{year}{1995}).

\bibitem[{\citenamefont{Frachebourg}(1999)}]{Frachebourg1999}
\bibinfo{author}{\bibfnamefont{L.}~\bibnamefont{Frachebourg}},
  \bibinfo{journal}{Phys. Rev. Lett.} \textbf{\bibinfo{volume}{82}},
  \bibinfo{pages}{1502} (\bibinfo{year}{1999}).

\bibitem[{\citenamefont{Frachebourg
  et~al.}(2000{\natexlab{a}})\citenamefont{Frachebourg, Martin, and
  Piasecki}}]{FrachebourgPiasecki2000}
\bibinfo{author}{\bibfnamefont{L.}~\bibnamefont{Frachebourg}},
  \bibinfo{author}{\bibfnamefont{P.~A.} \bibnamefont{Martin}},
  \bibnamefont{and} \bibinfo{author}{\bibfnamefont{J.}~\bibnamefont{Piasecki}},
  \bibinfo{journal}{Physica A} \textbf{\bibinfo{volume}{279}}
  (\bibinfo{year}{2000}{\natexlab{a}}).

\bibitem[{\citenamefont{Frachebourg
  et~al.}(2000{\natexlab{b}})\citenamefont{Frachebourg, Martin, and
  Piasecki}}]{LF00}
\bibinfo{author}{\bibfnamefont{L.}~\bibnamefont{Frachebourg}},
  \bibinfo{author}{\bibfnamefont{P.}~\bibnamefont{Martin}}, \bibnamefont{and}
  \bibinfo{author}{\bibfnamefont{J.}~\bibnamefont{Piasecki}},
  \bibinfo{journal}{Physica A} \textbf{\bibinfo{volume}{279}},
  \bibinfo{pages}{69} (\bibinfo{year}{2000}{\natexlab{b}}).

\bibitem[{\citenamefont{Valageas}(2009)}]{BA1D}
\bibinfo{author}{\bibfnamefont{P.}~\bibnamefont{Valageas}},
  \bibinfo{journal}{Physica A} \textbf{\bibinfo{volume}{388}},
  \bibinfo{pages}{1031} (\bibinfo{year}{2009}).

\bibitem[{\citenamefont{Trizac and Krapivsky}(2003)}]{Trizac:2003}
\bibinfo{author}{\bibfnamefont{E.}~\bibnamefont{Trizac}} \bibnamefont{and}
  \bibinfo{author}{\bibfnamefont{P.~L.} \bibnamefont{Krapivsky}},
  \bibinfo{journal}{Phys. Rev. Lett.} \textbf{\bibinfo{volume}{91}},
  \bibinfo{pages}{218302} (\bibinfo{year}{2003}).

\bibitem[{\citenamefont{Brilliantov and Spahn}(2006)}]{BrilliantovSpahn2006}
\bibinfo{author}{\bibfnamefont{N.~V.} \bibnamefont{Brilliantov}}
  \bibnamefont{and} \bibinfo{author}{\bibfnamefont{F.}~\bibnamefont{Spahn}},
  \bibinfo{journal}{Math. Comput. Simulation} \textbf{\bibinfo{volume}{72}},
  \bibinfo{pages}{93} (\bibinfo{year}{2006}).

\bibitem[{\citenamefont{Brilliantov et~al.}(2018)\citenamefont{Brilliantov,
  Formella, and Poeschel}}]{BFP2018}
\bibinfo{author}{\bibfnamefont{N.}~\bibnamefont{Brilliantov}},
  \bibinfo{author}{\bibfnamefont{A.}~\bibnamefont{Formella}}, \bibnamefont{and}
  \bibinfo{author}{\bibfnamefont{T.}~\bibnamefont{Poeschel}},
  \bibinfo{journal}{Nature Commun.} \textbf{\bibinfo{volume}{9}},
  \bibinfo{pages}{797} (\bibinfo{year}{2018}).

\bibitem[{\citenamefont{Midya and Das}(2017)}]{Das}
\bibinfo{author}{\bibfnamefont{J.}~\bibnamefont{Midya}} \bibnamefont{and}
  \bibinfo{author}{\bibfnamefont{S.~K.} \bibnamefont{Das}},
  \bibinfo{journal}{Phys. Rev. Lett.} \textbf{\bibinfo{volume}{118}},
  \bibinfo{pages}{165701} (\bibinfo{year}{2017}).

\bibitem[{\citenamefont{Paul and Das}(2018)}]{Das1}
\bibinfo{author}{\bibfnamefont{S.}~\bibnamefont{Paul}} \bibnamefont{and}
  \bibinfo{author}{\bibfnamefont{S.~K.} \bibnamefont{Das}},
  \bibinfo{journal}{Phys. Rev. E} \textbf{\bibinfo{volume}{97}},
  \bibinfo{pages}{032902} (\bibinfo{year}{2018}).

\bibitem[{\citenamefont{Singh and Mazza}(2019)}]{Mazza}
\bibinfo{author}{\bibfnamefont{C.}~\bibnamefont{Singh}} \bibnamefont{and}
  \bibinfo{author}{\bibfnamefont{M.~G.} \bibnamefont{Mazza}},
  \bibinfo{journal}{Sci. Reports} \textbf{\bibinfo{volume}{9}},
  \bibinfo{pages}{9049} (\bibinfo{year}{2019}).

\bibitem[{\citenamefont{Singh and Mazza}(2018)}]{Mazza1}
\bibinfo{author}{\bibfnamefont{C.}~\bibnamefont{Singh}} \bibnamefont{and}
  \bibinfo{author}{\bibfnamefont{M.~G.} \bibnamefont{Mazza}},
  \bibinfo{journal}{Phys. Rev. E} \textbf{\bibinfo{volume}{97}},
  \bibinfo{pages}{022904} (\bibinfo{year}{2018}).

\bibitem[{\citenamefont{Brilliantov et~al.}(2009)\citenamefont{Brilliantov,
  Bodrova, and Krapivsky}}]{BrilBodKrap2009}
\bibinfo{author}{\bibfnamefont{N.~V.} \bibnamefont{Brilliantov}},
  \bibinfo{author}{\bibfnamefont{A.~S.} \bibnamefont{Bodrova}},
  \bibnamefont{and} \bibinfo{author}{\bibfnamefont{P.~L.}
  \bibnamefont{Krapivsky}}, \bibinfo{journal}{J. Stat. Mech.}
  \textbf{\bibinfo{volume}{P06011}} (\bibinfo{year}{2009}).

\bibitem[{\citenamefont{Chapman and Cowling}(1970)}]{ChapmanCowling:1970}
\bibinfo{author}{\bibfnamefont{S.}~\bibnamefont{Chapman}} \bibnamefont{and}
  \bibinfo{author}{\bibfnamefont{T.~G.} \bibnamefont{Cowling}},
  \emph{\bibinfo{title}{The Mathematical Theory of Non-uniform Gases}}
  (\bibinfo{publisher}{Cambridge University Press}, \bibinfo{address}{New
  York}, \bibinfo{year}{1970}).

\bibitem[{\citenamefont{Brilliantov and
  P\"oschel}(2004)}]{BrilliantovPoeschelOUP}
\bibinfo{author}{\bibfnamefont{N.~V.} \bibnamefont{Brilliantov}}
  \bibnamefont{and}
  \bibinfo{author}{\bibfnamefont{T.}~\bibnamefont{P\"oschel}},
  \emph{\bibinfo{title}{Kinetic Theory of Granular Gases}}
  (\bibinfo{publisher}{Oxford University Press}, \bibinfo{address}{Oxford},
  \bibinfo{year}{2004}).

\bibitem[{\citenamefont{Connaughton et~al.}(2017)\citenamefont{Connaughton,
  Dutta, Rajesh, and Zaboronski}}]{connaughton2016universality}
\bibinfo{author}{\bibfnamefont{C.}~\bibnamefont{Connaughton}},
  \bibinfo{author}{\bibfnamefont{A.}~\bibnamefont{Dutta}},
  \bibinfo{author}{\bibfnamefont{R.}~\bibnamefont{Rajesh}}, \bibnamefont{and}
  \bibinfo{author}{\bibfnamefont{O.}~\bibnamefont{Zaboronski}},
  \bibinfo{journal}{Europhys. Lett.} \textbf{\bibinfo{volume}{117}},
  \bibinfo{pages}{10002} (\bibinfo{year}{2017}).

\bibitem[{\citenamefont{Connaughton et~al.}(2018)\citenamefont{Connaughton,
  Dutta, Rajesh, Siddharth, and Zaboronski}}]{colmPRE2018}
\bibinfo{author}{\bibfnamefont{C.}~\bibnamefont{Connaughton}},
  \bibinfo{author}{\bibfnamefont{A.}~\bibnamefont{Dutta}},
  \bibinfo{author}{\bibfnamefont{R.}~\bibnamefont{Rajesh}},
  \bibinfo{author}{\bibfnamefont{N.}~\bibnamefont{Siddharth}},
  \bibnamefont{and}
  \bibinfo{author}{\bibfnamefont{O.}~\bibnamefont{Zaboronski}},
  \bibinfo{journal}{Phys. Rev. E} \textbf{\bibinfo{volume}{97}},
  \bibinfo{pages}{022137} (\bibinfo{year}{2018}).

\bibitem[{\citenamefont{Meehan and Whittingham}(2015)}]{NOHEP}
\bibinfo{author}{\bibfnamefont{M.~T.} \bibnamefont{Meehan}} \bibnamefont{and}
  \bibinfo{author}{\bibfnamefont{I.~B.} \bibnamefont{Whittingham}},
  \bibinfo{journal}{JCAP} \textbf{\bibinfo{volume}{12}}, \bibinfo{pages}{011}
  (\bibinfo{year}{2015}).

\bibitem[{\citenamefont{Coskuner et~al.}(2019)\citenamefont{Coskuner,
  Grabowska, Knapen, and Zurek}}]{Zurek19}
\bibinfo{author}{\bibfnamefont{A.}~\bibnamefont{Coskuner}},
  \bibinfo{author}{\bibfnamefont{D.~M.} \bibnamefont{Grabowska}},
  \bibinfo{author}{\bibfnamefont{S.}~\bibnamefont{Knapen}}, \bibnamefont{and}
  \bibinfo{author}{\bibfnamefont{K.~M.} \bibnamefont{Zurek}},
  \bibinfo{journal}{Phys. Rev. D} \textbf{\bibinfo{volume}{100}},
  \bibinfo{pages}{035025} (\bibinfo{year}{2019}).

\bibitem[{\citenamefont{Bodrova et~al.}(2020)\citenamefont{Bodrova, Osinsky,
  and Brilliantov}}]{bodrova2020}
\bibinfo{author}{\bibfnamefont{A.}~\bibnamefont{Bodrova}},
  \bibinfo{author}{\bibfnamefont{A.}~\bibnamefont{Osinsky}}, \bibnamefont{and}
  \bibinfo{author}{\bibfnamefont{N.}~\bibnamefont{Brilliantov}},
  \bibinfo{journal}{Sci. Rep.} \textbf{\bibinfo{volume}{10}},
  \bibinfo{pages}{693} (\bibinfo{year}{2020}).

\bibitem[{\citenamefont{Osinsky et~al.}(2020)\citenamefont{Osinsky, Bodrova,
  and Brilliantov}}]{bodrovamol}
\bibinfo{author}{\bibfnamefont{A.}~\bibnamefont{Osinsky}},
  \bibinfo{author}{\bibfnamefont{A.}~\bibnamefont{Bodrova}}, \bibnamefont{and}
  \bibinfo{author}{\bibfnamefont{N.}~\bibnamefont{Brilliantov}},
  \bibinfo{journal}{Phys. Rev. E} \textbf{\bibinfo{volume}{101}},
  \bibinfo{pages}{022903} (\bibinfo{year}{2020}).

\bibitem[{\citenamefont{Osinsky}(2020)}]{gen-smol}
\bibinfo{author}{\bibfnamefont{A.}~\bibnamefont{Osinsky}}, \bibinfo{journal}{J.
  Comp. Phys.} \textbf{\bibinfo{volume}{422}}, \bibinfo{pages}{109764}
  (\bibinfo{year}{2020}).

\bibitem[{\citenamefont{Matveev et~al.}(2014)\citenamefont{Matveev,
  Tyrtyshnikov, Smirnov, and Brilliantov}}]{matveev}
\bibinfo{author}{\bibfnamefont{S.}~\bibnamefont{Matveev}},
  \bibinfo{author}{\bibfnamefont{E.}~\bibnamefont{Tyrtyshnikov}},
  \bibinfo{author}{\bibfnamefont{A.}~\bibnamefont{Smirnov}}, \bibnamefont{and}
  \bibinfo{author}{\bibfnamefont{N.}~\bibnamefont{Brilliantov}},
  \bibinfo{journal}{Vychisl. Methody Programm.} \textbf{\bibinfo{volume}{15}},
  \bibinfo{pages}{1} (\bibinfo{year}{2014}).

\bibitem[{\citenamefont{Matveev et~al.}(2017)\citenamefont{Matveev, Krapivsky,
  Smirnov, Tyrtyshnikov, and Brilliantov}}]{matveev1}
\bibinfo{author}{\bibfnamefont{S.~A.} \bibnamefont{Matveev}},
  \bibinfo{author}{\bibfnamefont{P.~L.} \bibnamefont{Krapivsky}},
  \bibinfo{author}{\bibfnamefont{A.~P.} \bibnamefont{Smirnov}},
  \bibinfo{author}{\bibfnamefont{E.~E.} \bibnamefont{Tyrtyshnikov}},
  \bibnamefont{and} \bibinfo{author}{\bibfnamefont{N.~V.}
  \bibnamefont{Brilliantov}}, \bibinfo{journal}{Phys. Rev. Lett.}
  \textbf{\bibinfo{volume}{119}}, \bibinfo{pages}{260601}
  (\bibinfo{year}{2017}).

\end{thebibliography}
\end{document}